\documentclass[aps,prd,twocolumn,groupedaddress,showpacs]{revtex4}
\usepackage{graphicx}
\usepackage{dcolumn}
\usepackage{bm}
\usepackage{amssymb,amsmath,latexsym,amsfonts}

\begin{document}

\title{Homogeneous one-dimensional Bose--Einstein Condensate in the Bogoliubov's Regime}

\author{El\'ias Castellanos}
\email{ecastellanos@mctp.mx.} \affiliation{Mesoamerican Centre for Theoretical Physics 
\\Universidad Aut\'onoma de Chiapas.
\\ Ciudad Universitaria, Carretera Zapata Km. 4, Real del Bosque (Ter\'an), 29040, Tuxtla Guti\'errez, Chiapas, M\'exico.}

\begin{abstract}
We analyze the corrections caused by finite size effects upon the ground state properties of a homogeneous one-dimensional Bose--Einstein condensate. We assume from the very beginning that the Bogoliubov's formalism is valid and consequently we show that in order to obtain a well defined ground state properties, finite size effects of the system must be taken into account. Indeed, the formalism described in the present work allows to recover the usual properties related to the ground state of a homogeneous one--dimensional Bose--Einstein condensate but corrected by finite size effects of the system. Finally, this scenario allows us to analyze the sensitivity of the system when the Bogoliubov's regime is valid and when finite size effects are present. These facts open the possibility to apply these ideas to more realistic scenarios, e.g., low--dimensional trapped Bose--Einstein condensates.
\end{abstract}

\maketitle

\section{Introduction}

The quantum nature of Bose--Einstein condensates has been arduously investigated over the last years from both, the experimental and the theoretical point of view  \cite{becs,Dalfovo}. Even more, due to its quantum nature and its high experimental precision Bose--Einstein condensates could be also used as relevant test tools in gravitational physics  \cite{eli,eli1}. One importan feature related to this phenomenon is the analysis and study of lower dimensional Bose--Einstein condensates, see for instance \cite{1D,1DEx} and references therein. Although, one--dimensional Bose--Einstein condensates can never be reached, it is possible to obtain quasi--one--dimensional Bose--Einstein condensates by using very extremely anisotropic traps, leading to the obtention of quasi--one--dimensional Bose--Einstein condensates in laboratories \cite{1DEx}. 

Here it must mentioned that one--dimensional Bose--Einstein condensates has some pathological behavior in the thermodynamic limit \cite{yuka1}. In order to make the condensation possible, finite size effects of the system are required and consequently, the ground state energy per particle contributions must be taken into account. 

In a serie of papers \cite{paper60} it was showed that a one-dimensional system of zero-range interacting bosons in a flat--bottom box is exactly integrable via Bethe anzats, for all values of the coupling strength. The properties of the ground state in a box provides an input for a unified beyond-mean-field treatment of trapped quasi-one-dimensional atomic gases. However, due to the peculiarities exposed in \cite{yuka1} for low dimensional systems, particularly for one--dimensional Bose--Einstein condensates, is interesting to analyze and to explore the behavior of these systems when finite size corrections are taken into account. In other words, it is important to analyze the sensitivity of these systems to finite size corrections. As mentioned above, in the works \cite{paper60} some properties of the one--dimensional Bose--Einstein condensate was analyzed, for instance the energy of the ground state, the pressure and consequently the corresponding speed of sound. Moreover, in such papers it was showed that for dense systems in one dimension the Bogoliubov's formalism is appropriate to describe the properties of the corresponding N--body ground state in a certain regime. The scenario analyzed in  \cite{paper60} has a one nontrivial coupling constant which is proportional to the strength of the interactions and inverse proportional to the density of particles. When this coupling constant is small enough, i.e., the system lies in regime of high densities, the Bogoliubov's theory is seen to be valid ``at leas as far as the energy of the ground state energy is concerned'' as was emphasized in the aforementioned works.

In this aim, we will explore the properties of the ground state of a one--dimensional Bose--Einstein condensate assuming that the Bogoliubov's formalism is valid from the very beginning. However, as we will show later in the manuscript, a new ingredient is required in order to obtain a well define ground state properties, i.e., finite size corrections must be taken into account. We will probe that if one assumes that Bogoliubov's regime is valid, then finite size effects of the system are also needed in order to obtain a well defined ground state energy, ground state pressure and consequently a well defined speed of sound. It is remarkable that finite size corrections of the system allows us to recover the properties of the ground state exposed in Refs.\,\cite{paper60}, like the ground state energy and consequently the corresponding speed of sound in the limit of the Bogoliubov's regime validity. However, these properties would be corrected by the finite size contributions of the system. These facts motivate the analysis of  the sensitivity of the one--dimensional system to finite size corrections. Clearly, the research exposed in the present manuscript must be extended to more realistic situations, i.e., trapped one--dimensional Bose--Einstein condensates. This feature will be presented elsewhere. In the present manuscript we will explore the properties of the ground state energy, assuming from the very beginning that the use of the Bogoliubov's formalism is correct. Furthermore, we will probe that if the finite size effects of the system are also taken into account the usual theory for these systems is recovered. 
\section{Ground state energy and Bogoliubov's tranformations}
In order to obtain the properties of the ground state energy of a one--dimensional condensate, let us propose the following N-body one--dimensional  Hamiltonian
\begin{eqnarray}
\hat{H}&=&
\sum_{p=0} \frac{p^2}{2m}\,\, \hat{a}_{p}^{\dagger}\hat{a}_{p}+\frac{U_{0_{1D}}}{2L}\sum_{p=0}\sum_{q=0}\sum_{r=0}\hat{a}_{q}^{\dagger}\hat{a}_{r}^{\dagger}
 \hat{a}_{q+p}\hat{a}_{r-p}, \,\,\,\,\,\,\,\,  \label{Ham1}
\end{eqnarray}
where $U_{0_{1D}}$ is the one dimensional self interaction parameter which describes the interactions within the condensate with $L$ the corresponding characteristic length (width) of the system. Also, we assume that the creation and annihilation operators satisfy the usual canonical commutation relations for bosons. Additionally, the case $p=0$ depicts the corresponding ground state level. Under these circumstances we assume that below the condensation temperature $N_{0}\approx N$ and $\sum_{p \not=0} N_{p} <<N$, being $N$ the total number of particles, $N_{p}$ the number of particles in the excited states, and $N_{0}$ the number of particles in the ground state. 

It is noteworthy to mention that one--dimensional Bose--Einstein condensates has some pathological behavior in the thermodynamic limit \cite{yuka1}. In order to make the condensation possible, finite size effects of the system are required and consequently, the ground state energy per particle contributions must be taken into account. In other words, we will assume that the ground state is not the level $p=0$. Instead of this fact, we will assume that the minimum value for the momentum $p$ is given by $p_{0}=\hbar /L$ \cite{yuka1} which is a consequence of the Heisenberg's uncertainty principle. In this scenario, we will prove that when the Bogoliubov's approximation is valid, we recover the results obtained in Refs.\,\cite{paper60}, but corrected by the finite size effects of the system, which is a remarkable result by itself.\\ 
Keeping terms up to second order in $\hat{a}_{0}$, and consistently, within the same order of the approximation we assume that $ \langle \hat{a}_{0}^{\dag} \hat{a}_{0} \rangle \approx  \langle \hat{a}_{0}^{2} \rangle \approx  \langle \hat{a_{0}^{\dag^{2}}}  \rangle \approx  N$. Thus, the Hamiltonian (\ref{Ham1}) can be re-expressed as follows
\begin{eqnarray}
\hat{H} &= &\frac{U_{0_{1D}}N^2}{2L}+ \frac{(\hbar/L)^{2}}{2m}N +
\sum_{p \not=p_0} \frac{p^{2}}{2m}+\frac{U_{0_{1D}}N}{L}\hat{a}_{p}^{\dagger}\hat{a}_{p} \nonumber\\ &+& \sum_{p \not=p_0}\frac{U_{0_{1D}} N}{2L}\Bigl[\hat{a}_{p}^{\dagger}\hat{a}_{-p}^{\dagger}
+ \hat{a}_{p}\hat{a}_{-p}\Bigr]. 
\label{Ham2}
\end{eqnarray}

The second right hand term of Eq.\,(\ref{Ham2}), contain contributions due to the finite size effects of the system.

At this point it is appropiated to mention that the Hamiltonian for a one-dimensional Bose-Einstein condensate with a $\delta$-interacting potential can be diagonalized via Bethe \emph{anzats} \cite{paper60}. In such a case, the energy per particle was deduced and is given by
\begin{equation}
\label{paper60}
E_{n}=\frac{\hbar^{2}}{2m}n^{2}e[g(n)],
\end{equation}
where $e[g(n)]$ is function of the parameter $g(n)=U_{0_{1D}}/n$, with $n$ the corresponding density of particles. From the expression (\ref{paper60}) two limits can be obtained
\begin{equation}
\label{LD}
E_{n} \approx \frac{\pi^{2} \hbar^{2}}{6m} n^{2},
\end{equation} 
\begin{equation}
\label{HD}
E_{n} \approx \frac{U_{0_{1D}}}{2}n,
\end{equation}
where Eq.\,(\ref{LD}) corresponds to the low density limit associated with the case of infinitely strong interactions, usually referred as a gas of impenetrable bosons or Tonks--Girardeau gas \cite{tonks}. Let us add that expression Eq.\,(\ref{LD}) formally coincides with the one for free fermions which suggest some kind of Bose--Fermi duality in one--dimensional systems \cite{tonks,FB}. Conversely, Eq.\,(\ref{HD}) depicts the high density limit. The high density limit corresponds to the Thomas--Fermi energy functional \cite{cigarro} (Clearly,  integrated over the corresponding volume times the number of particles) first introduced by Bogoliubov \cite{Bog}. We must mention that the in three--dimensional case the mean--field approximation is valid for low densities contrary to the one--dimensional case where it is valid at high densities. In other words, the Bogoliubov's theory should be correct for small values of $g(n)$. Thus, if we assume that the Bogoliubov's formalism is valid, i.e., the system lies in the high density limit then, without of loss of generality, we are capable to applied from the very beginning the pseudo--potential method in order to diagonalize our Hamiltonian (\ref{Ham3}). However, as we will see later in the manuscript, a novel ingredient is needed, i.e., also finite size effects are required for the system.

In order to  obtain the ground state energy associated with our system, let us diagonalize the Hamiltonian (\ref{Ham3}) by introducing, as usual within this approximation, the so-called Bogoliubov's transformations \cite{pathria,Ueda} which we assume valid also in one--dimension:
\begin{equation}
\hat{a}_{p}=
\frac{\hat{b}_{p} -
\alpha_p \hat{b}_{-p}^{\dagger}}{\sqrt{1-\alpha_p^2}},\label{Bog1} \hspace{0.5cm}\hat{a}_{p}^{\dagger}=
\frac{\hat{b}_{p}^{\dagger} -
\alpha_p\hat{b}_{-p}}{\sqrt{1-\alpha_p^2}},
\end{equation}
where $\alpha^{2}_{p}$ is the Bogoliubov's coefficient. The operators $\hat{b}_{p}^{\dagger}$ and $\hat{b}_{p}$ are also creation and annihilation operators. It is straightforward to show that these operators also obey the canonical commutation relations for bosons \cite{pathria}. 
Inserting the Bogoliubov transformations (\ref{Bog1}) into the Hamiltonian (\ref{Ham2}), we are capable to obtain the following diagonalized Hamiltonian
\begin{eqnarray}
\label{Ham3}
\hat{H} &=& \frac{U_{0_{1D}}N^2}{2L}+ \frac{(\hbar/L)^{2}N}{2m} + \sum_{p \not=p_0}\sqrt{\epsilon_{p}\Bigl(\epsilon_{p}+\frac{2U_{0_{1D}}N}{L}\Bigl)}\,\,
\hat{b}_{p}^{\dagger}\hat{b}_{p}\nonumber\\&-&\frac{1}{2}\sum_{p \not=p_0}\Bigg[\frac{U_{0_{1D}}N}{L} +\epsilon_{p}-\sqrt{\epsilon_{p}\Bigl(\epsilon_{p}+\frac{2U_{0_{1D}}N}{L}\Bigr)}\Bigg].
\end{eqnarray}
In the above expression we have defined $\epsilon_{p}=p^{2}/2m$, i.e., $\epsilon_{p}$ is the energy for a single free particle. Notice that the last summation in the Hamiltonian (\ref{Ham3}) would be divergent as $(U_{0_{1D}}N/L)^{2}/2\epsilon_{p}$, as can be seen by performing an expansion of the last term in equation (\ref{Ham3}) for large $p$. 
We assume that the pseudo-potential $U_{ps}(x)$ can be expressed, without of loss of generality in one dimension as $ U_{ps}(x)=U_{0_{1D}}\delta(x)d/d\, x(x *\_)$ where the $*$ represents the usual product and $\delta(x)$ is the Dirac Delta function. Thus, for all practical purposes the divergence can be expressed as $(1/p ^{2})$. This leads to $1/x$ in the configuration space, after the application of the Fourier transform. Additionally, as was pointed out in Ref.\,\cite{paper60}, if the Fourier transform of the pseudo potential is everywhere nonnegative, then the Bogoliubov's formalism is correct, which in fact corresponds to the case deduced here. Therefore, these facts reinforce our approach, i.e., when Bogoliubov's theory is valid, it could be useful to describe some properties caused by finite size effects of the system. Thus, even in this scenario, the action of the pseudo potential $U_{ps}$ removes the divergence.

Consequentely, our Hamiltonian is now given by
\begin{eqnarray}
\label{Ham301}
\hat{H} &=& \frac{U_{0_{1D}}N^2}{2L} + \frac{(\hbar/L)^{2}\,N}{2m} + \sum_{p \not=p_0}\sqrt{\epsilon_{p}\Bigl(\epsilon_{p}+\frac{2U_{0_{1D}}N}{L}\Bigl)}
\,\, \hat{b}_{p}^{\dagger}\hat{b}_{p}\,\,\, \nonumber\\&+&\sum_{p \not=p_0}\Bigg\{-\frac{1}{2}\Bigg[\frac{U_{0_{1D}}N}{L} +\epsilon_{p}
-\sqrt{\epsilon_{p}\Bigl(\epsilon_{p}+\frac{2U_{0_{1D}}N}{L}\Bigr)} \,\,\,\,\,\,\,\,\,\, \\\nonumber&-& \Bigl(\frac{U_{0_{1D}}N}{L}\Bigr)^2\frac{1}{2 \epsilon_{p}}\Bigg]\Bigg\}. 
\end{eqnarray}
In order to obtain the ground state energy, we replace the last summation in the above equation by an integration together with the following dimensionless variable $z^{2} = \epsilon_{p}L/U_{0_{1D}}N$. Then, the following expression associated with the ground state energy of the system is obtained   
\begin{eqnarray}
E_{0}&=& \frac{U_{0_{1D}}N^2}{2L} + \frac{(\hbar/L)^{2}}{2m}\,N \nonumber\\ &-&\frac{L\sqrt{2m}}{4\pi\hbar}\Bigl(\frac{U_{0_{1D}}N}{L}\Bigr)^{3/2}\int_{\gamma}^{\infty} f(z)\,dz,
\label{Psedo2}
\end{eqnarray}
where the function $f(z)$ is given by
\begin{equation}
f(z)=1+z^2-z\sqrt{2+z^2}-\frac{1}{2z^2}.
\end{equation}
Notice that we have taken into account that the lower limit in the integral is not zero. The lower limit that takes into account the energy of the ground state, and consequently, finite size effects of the system is given by the following expression
\begin{eqnarray}
\gamma^{2} =\Bigl(\frac{L}{U_{0_{1D}}N}\Bigr)\frac{(\hbar/L)^{2}}{2m}.
\label{Psedo33}
\end{eqnarray}
We must mention that if we take $\gamma=0$ (which corresponds to the case $p=0$) in the integral Eq.\,(\ref{Psedo2}) clearly becomes divergent, and the condensation is never reached at finite temperature in the thermodynamic limit. In other words, finite size effects of the system must be included in order to reach the condensation at finite temperature and consequently we recover the usual results deduced in Ref.\,\cite{paper60}.

On the other hand, from the $N$--body Hamiltonian (\ref{Ham301}), we are able to recognize the energy of the Bogoliubov's excitations 
\begin{equation}
\label{Ek}
E_{p} = \sqrt{\epsilon_{p}\Bigl(\epsilon_{p}+
 \frac{2U_{0_{1D}}N}{L}\Bigr)}. 
\end{equation}
The limit $p\rightarrow 0$, associated with the above expression, leads to the following dispersion relation
\begin{equation}
\label{qpart}
E_{p}=p\sqrt{\frac{U_{0_{1D}} N}{mL}},
\end{equation}
which is identical to that for a phonon, i.e., proportional to the momenta.

Conversely, at the high--energy limit, $p\rightarrow \infty$,  expression (\ref{Ek}) reads
\begin{equation}
\label{part}
E_{p}=\epsilon_{p}+\frac{U_{0_{1D}}N }{L},
\end{equation}
which basically is the single particle energy spectrum corrected by the mean field contributions.

The crossover between the phonon spectrum (\ref{qpart}) and the single particle behavior (\ref{part}) allows to define the corresponding healing length
\begin{equation}
\xi = \frac{1}{\sqrt{2mU_{0_{1D}}N/L\hbar^{2}}},
\label{HL0}
\end{equation}
in other words, we recover the usual results.

Finally, let us made some comments about the functional form of $U_{0_{1D}}$. We assume that the one dimensional parameter $U_{0_{1D}}$ is given by \cite{tesis}
\begin{equation}
U_{0_{1D}}=-\frac{2\hbar^{2}}{ma_{1D}},
\end{equation}
where $a_{1D}$ is the scattering length in one dimension. The above expression suggest that the one-dimensional scattering length is function of the three dimensional scattering length $a_{3D}$, as in trapped Bose-Einstein condensates. Thus, $a_{1D}$ seems to be
\begin{equation}
\label{a1d}
a_{1D}=-\frac{L^{2}}{2 a_{3D}}\Bigl(1-C\frac{a_{3D}}{L}\Bigr)\approx-\frac{L^{2}}{2 a_{3D}} ,
\end{equation}
where we have assumed that $a_{3D}$ is much smaller than $L$. In this approximation the one dimensional self--interaction parameter is given by 
\begin{equation}
U_{0_{1D}}\approx \frac{4\hbar^{2}a_{3D}}{mL^{2}}.
\end{equation}

\section{Speed of sound and finite size corrections}
On another matter, we are capable to obtain from Eq.\,(\ref{Psedo2}) the ground state energy $E_{0}$ associated with our one--dimensional N-body Hamiltonian 
\begin{eqnarray}
E_{0}&=& \frac{U_{0_{1D}}N^2}{2L} + \frac{(\hbar/L)^{2}}{2m}\,N \nonumber\\ &-&
\frac{L\sqrt{2m}}{4\pi\hbar}\Bigg(\frac{U_{0_{1D}}N}{L}\Bigg)^{3/2} \Bigg[\frac{2\sqrt{2}}{3}-\frac{1}{2 \gamma}-\gamma\Bigg].
\label{Psedo3}
\end{eqnarray}
Consequently, we are capable to obtain the corresponding speed of sound $c_{0}$ by using the following definitions
\begin{equation}
\label{SS}
c_{0}^{2}=-\frac{L^{2}}{Nm} \Bigl(\frac{\partial P_{0}}{\partial L}\Bigr)=c_{1}+c_{2},
\end{equation}
where $P_{0}=-(\partial E_{0}/\partial L )$ is the corresponding ground state pressure. In the previous expression we have defined 
\begin{eqnarray}
c_{1}& = & \frac{24 a_{3D} \hbar^2 N}{L^3 m^2}-\frac{42 a_{3D} \hbar \sqrt{\frac{a_{3D} \hbar^2 N}{L^3 m}}}{\pi  L^2 m^{3/2}}, \nonumber
\end{eqnarray}
\begin{eqnarray}
 \nonumber\\&&c_{2}  =  \frac{396\,a_{3D}^{2} \hbar N \sqrt{\frac{a_{3D} \hbar^2 N}{\sqrt{2} L^3 m}}}{\pi  L^3 m^{3/2}}+\frac{35 \hbar \sqrt{\frac{a_{3D} \hbar^2 N}{L^3 m}}}{8 \sqrt{2} \pi  L m^{3/2} N}+\frac{3 \hbar^2}{L^2 m^2},\,\,\,\,\,\,\,\,\,  \nonumber 
\end{eqnarray}

Here we also recover the results obtained in Refs.\,\cite{paper60} concerning to the speed of sound plus corrections due to the finite size effects of the condensate, which are encoded in $c_{2}$. 

In order to analyze the sensitivity of our system to the corrections caused by the finite size effects of the system, we will take fiducial laboratory conditions over the parameters related  to the model as, $N\sim 10^{4}$ particles, $a_{3D} \sim10^{-9}$ m, and $m\sim 10^{-26}$ kg,  \cite{Dalfovo}. Together with $L\sim 10^{-3}$m. In our case we assume that the corresponding speed of sound is of order $3\times10^{-3}$\, m$s^{-1}$, i.e., the same order of magnitud as in a three--dimensional Bose-Einstein condensate \cite{Andrews}. It is clear that although, from the experimental point of view there is no a real one--dimensional condensate, it is possible to construct a quasi-one-dimensional condensate just by using extremely anisotropic traps \cite{aniso}. Then, as a first step, the use of this experimental accuracy is justified in order to analyze the sensitivity of the system to finite size corrections.

The experimental parameters mentioned above allows us to estimate the relative shift caused by finite size effects of the system, which can be inferred up to 
\begin{equation}
\frac{\Delta c_{0}}{c_{1}} \equiv \frac{c_{0}-c_{1}}{c_{1}}\approx  10^{-5}\,.
\label{sh}
\end{equation}
where $c_{1}$ is the usual value without finite size effects corrections.
The relative shift in the speed of sound is of order $\sim 10^{-5}$\,m$s^{-1}$, i.e., at least two orders of magnitude smaller than the typical value $10^{-3}\,$m$s^{-1}$, in typical experimental conditions. However, if we assume that the scattering length is of order $10^{-6}$ then, the relative shift expression (\ref{sh}) can be inferred up to $10^{-2}$. In other words, the sensitivity to finite size corrections for this \emph{toy model} is more viable when the strength of interactions is considerable.

The problem of large values of $a_{3D}$ can be solved, in principle, just tuning the interaction coupling by Feshbach resonances to different values of the scattering length $a_{3D}$, and then, modulate the contribution of interactions on the speed of sound below the \emph{finite size} induced shift for a sufficient dense systems. Nevertheless, these facts could affect the stability of the condensate and can give rise to technical difficulties. Furthermore, this scenario must be extended to a more realistic situation, i.e., one-dimensional trapped Bose-Einstein condensates.

\section{Conclusions}
We have analyzed a homogeneous one--dimensional Bose--Einstein condensate assuming that the Bogoliubov's formalism is valid. Even more, in order to obtain the usual theory related to these systems and also obtain well defined ground state properties, finite size effects must be taken into account. We have calculated the corrections on the ground state energy and consequently the corrections in the corresponding speed of sound, caused by finite size effects of the corresponding system. These facts leads to a shift in the associated speed of sound of order $10^{-5}$ in typical laboratory conditions. 

We must empathize that this model could be interpreted as a some kind of \emph{toy model}, since as far as we know, homogeneous condensates are not possible in experiments, i.e., there is no condensates in a box. Realistic condensates are bosonic systems trapped in magnetic traps among others \cite{yuka1,Dalfovo}. Nevertheless, the model analyzed here open up the opportunity to extend these ideas to more realistic systems, i.e., trapped condensates in one or two dimensions, in order to analyze the sensitivity of these systems to finite size effects. Furthermore,  the results exposed in the present work could be also used as test tools in some lines of gravitational physics from the phenomenological point of view \cite{elias2}.  

Finally, we stress that the many-body contributions in a Bose-Einstein condensate open the possibility of planning specific scenarios that could be used, in principle, to analyze the sensitivity of these low--dimensional systems to finite size effects. 

\begin{acknowledgments}
E.C.  acknowledges MCTP/UNACH for financial support.
\end{acknowledgments}

\end{document}